\documentclass[twocolumn,amsmath,superscriptaddress,amssymb,aps,prl,showpacs]{revtex4-1}

\usepackage{graphicx,txfonts}
\usepackage[colorlinks, linkcolor=blue, citecolor=blue, urlcolor=blue, breaklinks=true]{hyperref}

\newcommand{\eref}[1]{Eq.~(\ref{#1})}
\newcommand{\erefs}[1]{Eqs.~(\ref{#1})}
\newcommand{\fref}[1]{Fig.~\ref{#1}}

\DeclareMathOperator{\Tr}{Tr}

\begin{document}

\title{Thermodynamics of trajectories of a quantum harmonic oscillator coupled to $N$ baths}

\author{Simon Pigeon}
\email[Corresponding author:\ ]{s.pigeon@qub.ac.uk}
\affiliation{Centre for Theoretical Atomic, Molecular and Optical Physics, School of Mathematics and Physics, Queen's University Belfast, Belfast BT7\,1NN, United Kingdom}
\author{Lorenzo Fusco}
\affiliation{Centre for Theoretical Atomic, Molecular and Optical Physics, School of Mathematics and Physics, Queen's University Belfast, Belfast BT7\,1NN, United Kingdom}
\author{Andr\'e Xuereb}
\affiliation{Department of Physics, University of Malta, Msida MSD2080, Malta}
\affiliation{Centre for Theoretical Atomic, Molecular and Optical Physics, School of Mathematics and Physics, Queen's University Belfast, Belfast BT7\,1NN, United Kingdom}
\author{Gabriele De Chiara}
\affiliation{Centre for Theoretical Atomic, Molecular and Optical Physics, School of Mathematics and Physics, Queen's University Belfast, Belfast BT7\,1NN, United Kingdom}
\author{Mauro Paternostro}
\affiliation{Centre for Theoretical Atomic, Molecular and Optical Physics, School of Mathematics and Physics, Queen's University Belfast, Belfast BT7\,1NN, United Kingdom}

\date{\today}

\begin{abstract}
We undertake a thorough analysis of the thermodynamics of the trajectories followed by a quantum harmonic oscillator coupled to $N$ dissipative baths by using a new approach to large-deviation theory inspired by phase-space quantum optics. As an illustrative example, we study the archetypal case of a harmonic oscillator coupled to two thermal baths, allowing for a comparison with the analogous classical result. In the low-temperature limit, we find a significant quantum suppression in the rate of work exchanged between the system and each bath. We further show how the presented method is capable of giving analytical results even for the case of a driven harmonic oscillator. Based on that result, we analyse the laser cooling of the motion of a trapped ion or optomechanical system, illustrating how the emission statistics can be controllably altered by the driving force. 
\end{abstract}


\maketitle

Finding a concise description of the dynamics of a quantum system connected to an environment is one of the challenges of modern quantum physics. Even when the unitary evolution of a quantum system is well-known, its open dynamics is often less clear~\cite{Bellomo2007}; the full description of the exchange of excitations between a system and its environment would find much application in both experimental and theoretical analysis of open quantum systems. The primary tools one has access to in analysing such dynamics efficiently are input-output theory~\cite{Gardiner2010} and full counting statistics~\cite{Belzig2001}. More recently, a promising approach came to light based on large-deviation theory~\cite{Garrahan2010}.

In the context of statistical physics, large-deviation theory has led to the development of a powerful tool to characterize the dynamic and thermodynamic behavior of non-equilibrium classical systems~\cite{Lecomte2005}. Based on the long-time limit of the probability distribution associated with the trajectories of a particular observable, this method is nowadays used as a benchmark to define thermodynamic quantities and relations in classical non-equilibrium systems~\cite{Marconi2008,Touchette2009}. Following the proposal put forward in Ref.~\cite{Garrahan2010} to extend this method to the quantum regime, several different problems have been addressed, proving the efficacy of this approach in shedding new light on the dynamics of exchange between a quantum system and its environment~\cite{Ates2012,Genway2014,Pigeon2014}. 

Despite being at the heart of a rather effective method, the calculation of the large-deviation function itself---which encodes all the relevant information about a chosen counting process---often faces practical difficulties, in both classical and quantum cases. Indeed, in most cases the analytical calculation of the large-deviation function remains out of reach, while the numerical estimation of such function is strongly affected by the size of the phase or Hilbert space of the system.  

In this paper we consider a paradigmatic system in quantum mechanics, a quantum harmonic oscillator connected to $N$ arbitrary baths whose dynamics is governed by a master equation in Lindblad form. This system is a fundamental building block of quantum optics and is used to describe a large variety of quantum degrees of freedom, including the motion of trapped ions and molecules, cavity and circuit quantum electrodynamic systems, and many-body systems. One of the key results of this paper is an analytical expression for the large-deviation function of this frequently-encountered infinite-dimensional Hilbert space problem (i.e., a continuous-variable system). In the case where the harmonic oscillator is coupled to two thermal baths, we compare our results to the classical case, showing perfect agreement at high temperatures and an unexpected quantum suppression at low temperatures. Following this, we analyze a driven harmonic oscillator, again presenting analytical results for the large-deviation function. Far from being an exclusively descriptive approach, we show how to engineer the output of a quantum harmonic oscillator to read physically meaningful internal quantities. As an example of the practical application of the presented method we will consider a trapped ion or optomechanical system and see how the large-deviation function gives access to internal degrees of freedom through the outgoing flux of quanta, and how driving can significantly enhance this flux.

\noindent\emph{Model.}---We consider a quantum harmonic oscillator coupled to $N$ baths; the system Hamiltonian is defined as $H= \omega({a}^\dagger {a}+\tfrac12 )$ with $\omega$ the harmonic oscillator frequency (we use units such that $\hbar=1$). Under the Born--Markov approximation, the coupling between the system and the $i$\textsuperscript{th} bath ($1\le i\le N$) is modeled through the superoperator
\begin{equation}
\mathcal{L}_i[\bullet] = \bar{\Gamma}_i \bigl(2{a}^{\dagger}\bullet{a}-\bigl\{ \bullet,{a}{a}^{\dagger}\bigr\} \bigr)
+ \Gamma_i\bigl(2{a}\bullet{a}^{\dagger}-\bigl\{ \bullet,{a}^{\dagger}{a}\bigr\} \bigr), \label{ll}
\end{equation} 
allowing the exchange of quanta both to the bath (with a rate $\Gamma_i\geq0$) and from it ($\bar\Gamma_i\geq0$). The dynamics of such a system will obey the master equation $\partial_t \rho = \mathcal{W}[\rho]$ where $\rho$ is the density matrix associated to the harmonic oscillator and $\mathcal{W}[\bullet]=-i[ H,\bullet]+\sum_{i=1}^N\mathcal{L}_i[\bullet]$ is the superoperator governing the open dynamics.

In order to describe the dynamical behavior of the system we will focus on a counting process $K$ associated with the \emph{net} number of quanta leaving the system to bath $1$, which we refer to as the reference bath. We can unravel the master equation of the reduced density matrix by projecting it onto a particular number of quanta, i.e., $\partial_t \rho= P^K\mathcal{W}[\rho]$ where $P^K$ is a projector over trajectories counting $K$ net exchanged quanta. Thus, $p_K(t)=\Tr\{P^K\rho(t)\}$ represents the probability of observing such a trajectory. The moment-generating function associated to $p_K(t)$ is $Z(t,s)=\sum_{K=0}^\infty e^{-sK}p_K(t)=\Tr \{ \rho_s(t) \}$~\cite{Garrahan2010}, with $\rho_s(t)=\sum_{K=0}^\infty e^{-sK}\rho^K(t)$, where $s$ is the \emph{bias parameter}. The biased density operator $\rho_s(t)$ evolves according to the modified master equation $\partial_t\rho_s=(\mathcal{W}+\mathcal{L}_s)[\rho_s]$, where 
\begin{equation}
\mathcal{L}_{s}[\bullet]= 2\Gamma_1(e^{-s}-1){a}\bullet{a}^{\dagger}+2\bar\Gamma_1(e^{s}-1){a}^{\dagger}\bullet{a}.\label{ls}
\end{equation}
In the long-time limit, large-deviation theory applies and we can write $Z(t,s)\to e^{t\theta(s)}$, where the large-deviation function $\theta(s)$ represents the system's dynamical free energy~\cite{Garrahan2010}. Consequently, we have $\theta(s)=\lim_{t\to\infty}\ln\bigl(\Tr\{\rho_{s}\}\bigr)/t$.

We define the symmetrically-ordered characteristic function associated to $\rho_s(t)$, i.e., $\chi(\beta)=\text{Tr}\bigl\{ \exp(i \beta{a}^{\dagger}-i \beta^*{a})\rho_{s}\bigr\} 
 $~\cite{Gardiner2010} with $\beta\in{\mathbb C}$ and $\beta^*$ its complex conjugate. This leads to the Fokker--Planck equation $\partial_{t}\chi(\beta)=\mathcal{X}[\chi(\beta)]$, where
\begin{multline}
\label{xi}
\mathcal{X}[\bullet]=\bigl[\bigl(i \omega+\tfrac{\Delta_-}{2} \bigr)\beta^{*}\partial_{\beta^{*}}-\bigl(i\omega-\tfrac{\Delta_-}{2}\bigr)\beta\partial_{\beta}-\tfrac{\Delta_+}{2}|\beta|^2\\
	-4f_+(s)\bigl( \partial_{\beta}\partial_{\beta^{*}}+\tfrac{1}{4} |\beta|^2 \bigr)	-2f_-(s)\bigl( \beta^{*}\partial_{\beta^{*}}+\beta\partial_{\beta}+1\bigr)\bigr]\bullet,
 \end{multline}
$\Delta_\pm=\sum_{i=1}^N(\Gamma_i\pm \bar\Gamma_i)$, and $f_{\pm}(s)=\tfrac{1}{2}\bigl[\Gamma_1(e^{-s}-1)\pm\bar{\Gamma_1}(e^{s}-1)\bigr]$. A key feature of the present system and its associated dynamics is that, owing to the quadratic nature of the operators being involved, the Gaussian nature of the characteristic function is ensured, even including the superoperator (\ref{ls}) in the dynamics. Consequently, we will consider a Gaussian ansatz
\begin{align}
\chi(p, q)=A(t)\exp\left[i \mathbf{u'\boldsymbol\mu}(t)-\tfrac{1}{2}\mathbf{u'\Sigma}(t)\mathbf{u}\right]
,\label{ansatz}
 \end{align}
where $\mathbf{u}=(p,q)'$, $\boldsymbol\mu (t) = (x(t),y(t))'$ is vector of first moments, $\mathbf{\Sigma}(t)$ is the covariance matrix, and $A(t)$ is an amplitude which arises due to the trace non-preserving character of the superoperator in \eref{ls}. Normalization of $\rho_s(t)$, i.e., $A(t)=1$, is recovered for $s=0$. In writing $\chi(\beta)$ we introduced the quadrature variables $q=(\beta+\beta^*)/\!\sqrt{2}$ and $p=i(\beta^*-\beta)/\!\sqrt{2}$. Using the completeness of the coherent states, we can write $\rho_s=(1/\pi)\int\chi(\beta)D^\dag(\beta)d^2\beta$~\cite{Cahill1969} with $D(\beta)=\exp[\beta b^\dag-\beta^* b]$ the displacement operator. Finally, noting that $\Tr\bigl\{D(\beta)\bigr\}=\pi\delta^2(\beta)$, we have $\Tr\bigl\{\rho_{s}\bigr\}=A(t)$. The procedure used to define $\rho_s(t)$ through the introduction of the superoperator (\ref{ls}) has the effect of encoding the information relating to the counting process in the trace of $\rho_s(t)$. Consequently, determining $A(t)$ is sufficient to obtain the large-deviation function $\theta(s)$. 

Inserting \eref{ansatz} into \eref{xi} and identifying the different moments of $p$ and $q$ we can determine the evolution of the different parameters entering $\chi(\beta)$. Starting from an initial thermal state, $x(0)=y(0)=\sigma_{xy}(0)=0$, we can deduce that $x(t)=y(t)=\sigma_{xy}(t)=0$ for all $t$. This leads to a symmetric solution, such that $\sigma_x(t)=\sigma_y(t)\equiv\sigma(t)$ for all $t$. The evolution equations of the Gaussian parameters then reduce to two:
\begin{align}
\label{sigmaA}
\partial_{t}\sigma(t)&=-2\bigl[\Delta_-+2f_{-}(s)\bigr]\sigma(t)+2\Delta_++2f_{+}(s)\bigl[\sigma(t)^{2}+1\bigr],\nonumber\\
\partial_{t}A(t)&=2\bigl[f_{+}(s)\sigma(t)-f_{-}(s)\bigr]A(t).
\end{align}
In this case, the large-deviation function can be written as 
\begin{equation}
\theta(s) =	2\lim_{t\to \infty}\frac{1}{t}\int^{t}\bigl[f_+(s)  \sigma(\tau)-f_-(s)\bigr]\rm{d}\tau \label{theta0}.
\end{equation}
Notice that the effects of the chosen initial conditions and any transients are of little interest in our case, since they disappear in the long-time limit. By the uniqueness of the steady state, any initial condition will evolve to the state described by $\theta(s)$.

In order for the obtained state to be physically admissible, $\sigma(t)$ must be real and positive. To fulfil this criterion we require that $\Gamma_1\ge\bar\Gamma_1$ and $s^{-}\leq s\leq s^{+}
 $, where
\begin{equation}
e^{s^{\pm}}=\tfrac{1}{A}\bigl(B\pm\sqrt{B^{2}-AC}\bigr),
\label{ss}
\end{equation}
$A=4\bar{\Gamma}_1(\Delta_++\Delta_--2\Gamma_1)$, $B=2\bar{\Gamma}_1(\Delta_++\Delta_-)+2\Gamma_1(\Delta_+-\Delta_--4\bar\Gamma_1)+\Delta_-^{2}$, and $C=4\Gamma_1(\Delta_--\Delta_++2\bar{\Gamma}_1)$. This ensures the existence of a steady state, and thus stationary values of both $\sigma(t)$ and $A(t)$, so that we can legitimately use large-deviation theory~\cite{Touchette2009}. After a straightforward calculation we find the large-deviation function
\begin{equation}
\theta(s)=\Delta_- - \sqrt{[\Delta_-+2f_{-}(s)]^{2}-4f_{+}(s)[\Delta_++f_{+}(s)]},
\label{theta}
\end{equation}
which is such that $\theta(0)=0$, as required by the normalization of $\rho(t)$.

Having at hand the large-deviation function, we can obtain access to key figures of merit of the system. The most immediate figure is the \emph{activity} $k(0)=-\partial_s \theta(s)\vert_{s=0}$, which represents the mean rate of excitations exchanged between the system and the reference bath. In our case, it takes the explicit form $k(0)=\bigl(\Gamma_1-\bar{\Gamma}_1\bigr)\Delta_+/\Delta_--\bigl(\Gamma_1+\bar{\Gamma}_1\bigr)$. A second quantity of much relevance to quantum optics that is directly accessible from $\theta(s)$ is the Mandel $Q$-factor~\cite{Mandel1979}, $Q(0)=-\partial^2_{s} \theta(s)/\partial_s \theta(s)\vert_{s=0}-1$. The $Q$-factor is related to the variance of the number of exchanged quanta, and is therefore useful in characterising counting statistics. In the specific case where $\bar\Gamma_1=0$ (e.g., when the bath in question is thermal and at zero temperature), the Mandel $Q$-factor will coincide with the variance of emitted quanta to the reference bath and be expressed as $Q(0)=\Gamma(\Delta_+-\Delta_-)/\Delta_-^{2}$. For physical reasons we need to have $\Delta_+\ge \Delta_-$; we can thus conclude that no anti-bunching [$Q(0)<0$] can be obtained with the scenario in question~\cite{NoteQfactor}: A quantum harmonic oscillator coupled to any number of baths will not exhibit anti-bunching in the excitations it emits to any one of those baths.

\begin{figure}[t]
 \includegraphics[width=0.4\textwidth]{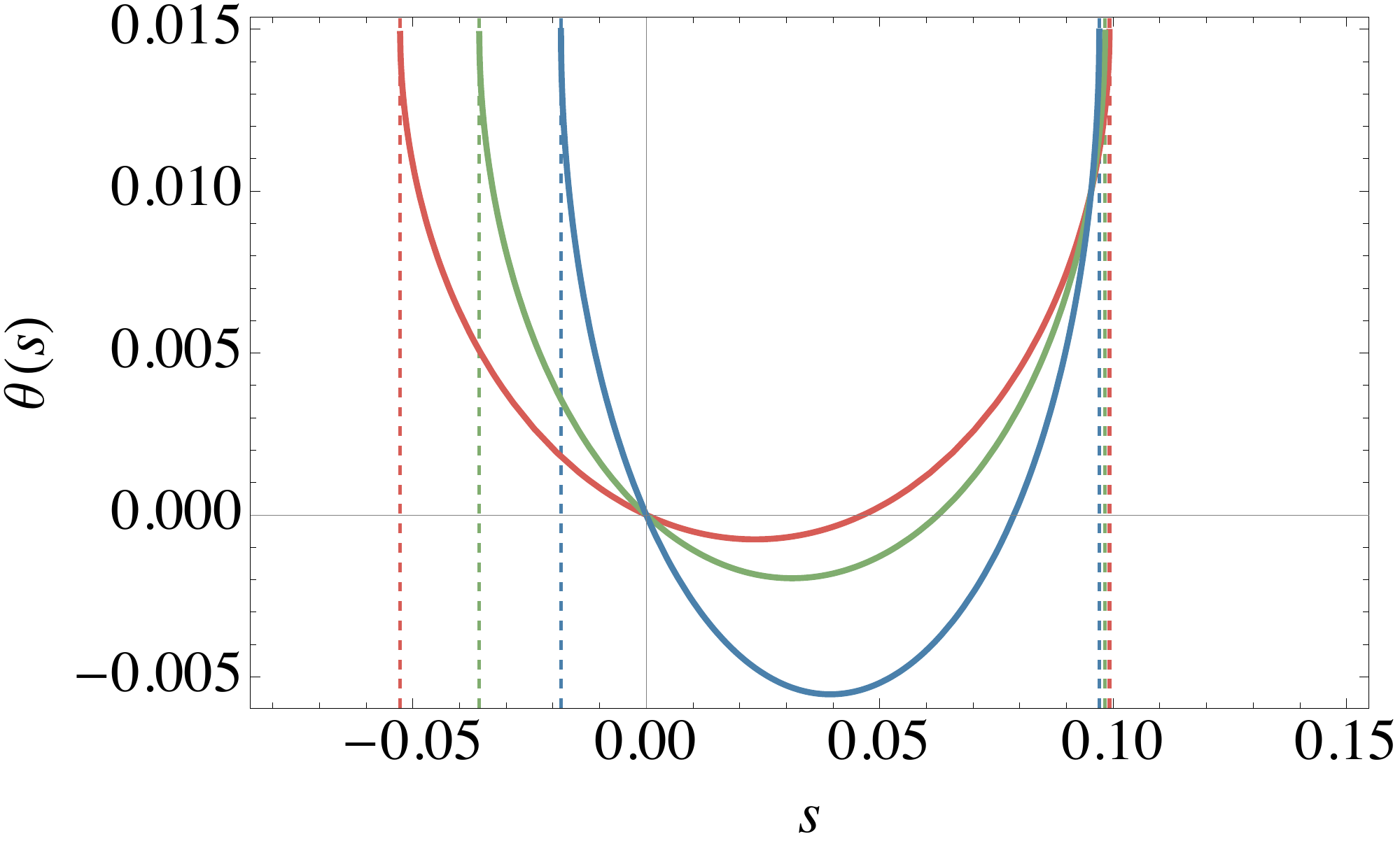}
\caption{(Color online) Large-deviation function $\theta(s)$ for different temperatures of the second thermal bath. From top to bottom ($0<s\lesssim0.1$), we have $n_2=20$ (red), $n_2=30$ (green), and $n_2=60$ (blue). The vertical dashed lines highlight the branch points for each curve. ($n_1=10$, $\gamma_1=\gamma_2/2=0.01$.)}
  \label{fig:thermostat}
\end{figure}

\noindent\emph{Thermal-bath case.}---Let us consider the explicit situation of two thermal baths, where $\Gamma_i\equiv\gamma_i(n_i+1)/2$ and $\bar\Gamma_i\equiv\gamma_in_i/2$, with $n_i$ the mean number of excitations in bath $i=1,2$ and $\gamma_i$ the coupling strength between the system and the bath. For $N>2$ baths, we may group the baths with $i\ge2$ into one ``superbath''; the situation we consider here is thus general. We show in \fref{fig:thermostat} the corresponding large-deviation function $\theta(s)$, as defined in \eref{theta}, for different temperatures $n_2$ of the second bath. We can see that increasing this temperature tends to increase the curvature of $\theta(s)$ and brings about an associated increase of the activity $k(0)$. The curves demonstrate the expected Gallavotti--Cohen symmetry property~\cite{Pigeon2014,Lebowitz2011} of $\theta(s)$, i.e., $\theta(s)$ is symmetric about some $s=s_0$, and features two branch points, at $s_+$ and $s_-$; these branch points are related to exponential tails of the probability distribution $p_K(t)$, where the large-deviation ansatz fails~\cite{Visco2006}.

\begin{figure}[t]
 \includegraphics[width=0.35\textwidth]{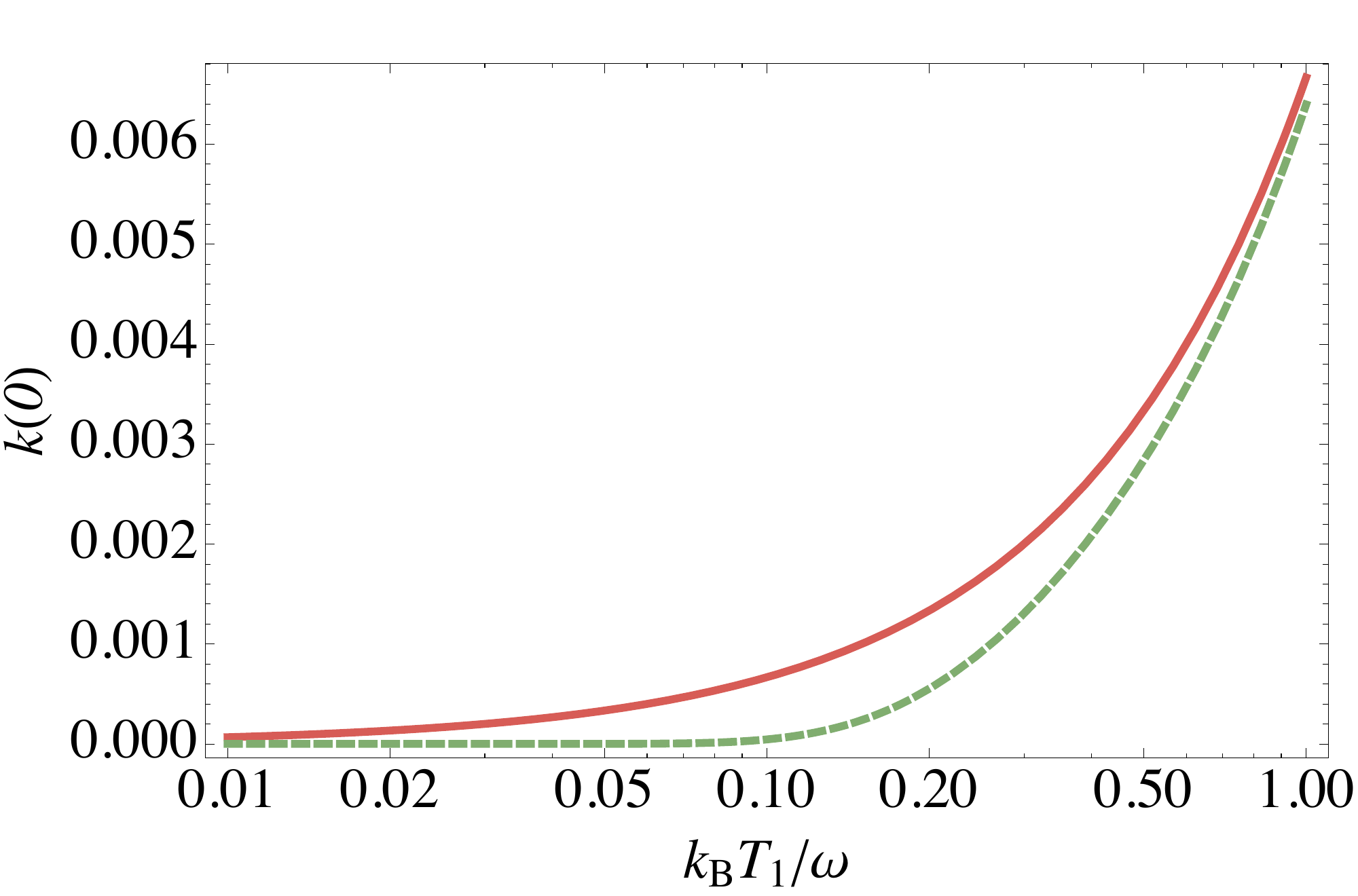}
\caption{(Color online) Mean rate of excitations exchanged [i.e., activity $k(0)$] between the harmonic oscillator and its reference bath as a function of the bath temperature $T_1$. The classical case is illustrated by the red solid curve and the quantum case by the green dashed curve. ($T_2=2T_1$, $\gamma_1=\gamma_2/2=0.01$.)}
  \label{fig:bench}
\end{figure}

For the classical analog of this system, the associated large-deviation function has been derived analytically through different methods~\cite{Visco2006,Fogedby2011}; the resulting function also exhibits two branch points and the Gallavotti--Cohen symmetry. To compare the two results quantitatively, we note that the number of excitations $n_i$ in each bath can be related to a temperature $T_i$ as $n_i=1/\bigl(e^{\hbar\omega/k_\mathrm{B}T_i}-1\bigr)$~\cite{Gardiner2010}. In the high-temperature limit, $n_i\sim T_i$ and we obtain a perfect match between the classical large-deviation function~\cite{Visco2006,Fogedby2011} and its quantum counterpart \eref{theta}. Turning to the opposite limit, where $T_i\to 0$, we find that a prominent difference appears: There is a significant suppression of activity in the quantum case. These features are shown in \fref{fig:bench}, where we plot the net rate of excitations exchanged between the system and bath $1$ [i.e., the activity $k(0)$] as a function of the temperature $T_1$ of this bath, for both classical and quantum harmonic oscillators. At high temperatures, the classical (full red line) and quantum (green dashed) results converge, whilst at low temperatures the two differ markedly. The net number of excitations exchanged, which can be related to the average work done by the harmonic oscillator on the reference bath, is smaller in the quantum case than in the classical one; we conjecture that this is due to the discrete nature of the excitation-exchange process in the quantum case. The approach developed here allows us to investigate these features without the limitations imposed by a numerical approach based on a truncated Hilbert space.

\noindent\emph{Driven harmonic oscillator.}---We now consider the case of a driven harmonic oscillator. The Hamiltonian will be modified to include a driving term $\mathcal{F}(t) ({a}^\dagger +{a})$, where $\mathcal{F}(t)$ represents the time-dependent amplitude of the driving force. The calculation in this case proceeds similarly to the preceding one, the difference being that the dynamical equation for the characteristic function, $\mathcal{X}[\bullet]$, will now include the driving-dependent term $4\omega\mathcal{F}(t)\bullet$. Under these conditions, the set of differential equations for the Gaussian parameters can no longer be reduced to the simple form \eref{sigmaA}, since we have
\begin{align}
\label{x&y}
\partial_{t}x(t)&=-\omega y(t)+\bigl[2f_{+}(s)\sigma(t)-2f_{-}(s)-\Delta_-\bigr]x(t),\\
\partial_{t}y(t)&= \omega [x(t)+4\mathcal{F}(t)]+\bigl[2f_{+}(s)\sigma(t)-2f_{-}(s)-\Delta_-\bigr]y(t).\nonumber
\end{align}
Furthermore, the second equation in \erefs{sigmaA} is modified to
\begin{equation}
\partial_{t}A(t)=\bigl\{f_{+}(s)\bigl[2\sigma(t)+x^2(t)+y^2(t)\bigr]-2f_{-}(s)\bigr\}A(t).
\label{aa}
\end{equation}
Consequently, we can write $\theta(s)=\theta_{\text{osc}}(s)+\theta_{\text{d}}(s)$, where $\theta_{\text{osc}}(s)$ stands for the large-deviation function of the bare harmonic oscillator as defined in \eref{theta}, and $\theta_{\text{d}}(s)$ the contribution coming from the driving. Solving \erefs{x&y} analytically is possible. As was done previously, we consider the long-time limit, where any transient effects are discarded. Consequently, we may simply consider the steady state $\sigma_\mathrm{st}$ of $\sigma(t)$ to solve \erefs{x&y}. We find, for $s_-\le s\le s_+$ and independently of $\mathcal{F}(t)$,
\begin{equation}
 \sigma_\mathrm{st}= \bigl[\Delta_-+2 f_{-}(s) -\Sigma(s)\bigr]/\bigl[2f_{+}(s)\bigr]
 \end{equation}
with $\Sigma(s)=\sqrt{[\Delta_- +2 f_{-}(s)]^2-4 f_{+}(s)[\Delta_+ +f_{+}(s)]}$. In order to proceed further, we must specify the form of the driving force $\mathcal{F}(t)$; we shall now consider two cases, (i)~constant driving, and (ii)~periodic driving. In the former case, we take $\mathcal{F}(t)=\mathcal{F}$, and the driving contribution to $\theta(s)$ is
\begin{equation}
\theta_{\text{d}}(s)= \frac{16\mathcal{F}^2\omega^2}{\Sigma(s)^{2}+\omega^2} f_{+}(s)
\label{thetadc}.
\end{equation}
In the latter, we take $\mathcal{F}(t)=\mathcal{F}\cos[(\omega +\delta )t]$ and find 
\begin{equation}
\theta_{\text{d}}(s)=\frac{8\mathcal{F}^2\omega^{2}}{\Sigma(s)^{2}+\delta^{2}}\left[1- \frac{2\omega\left(\omega+\delta\right)}{\Sigma(s)^{2}+\left(2\omega+\delta\right)^{2}}\right]f_{+}(s) \label{thetada}
\end{equation}
for $\delta\neq-\omega$. In the special case $\delta=-\omega$, the driving is no longer periodic and \eref{thetadc} must be used; continuity of $\theta_\mathrm{d}(s)$ at $\delta=-\omega$ fails because of the interplay between the long-time limit and the time-dependence of the driving~\cite{NoteLimits}. However, from both results we see that the large-deviation function scales as the square of the excitation amplitude, leading to the conclusion that the activity will scale quadratically with the excitation amplitude. 

In order to see how these results apply to practical situations, we consider now the cooling of the motional degree of freedom of a trapped ion, or an equivalent optomechanical system, and use the large-deviation function to interpret the resulting statistics of excitations exchanged.

\begin{figure}[t]
 \includegraphics[width=0.45\textwidth]{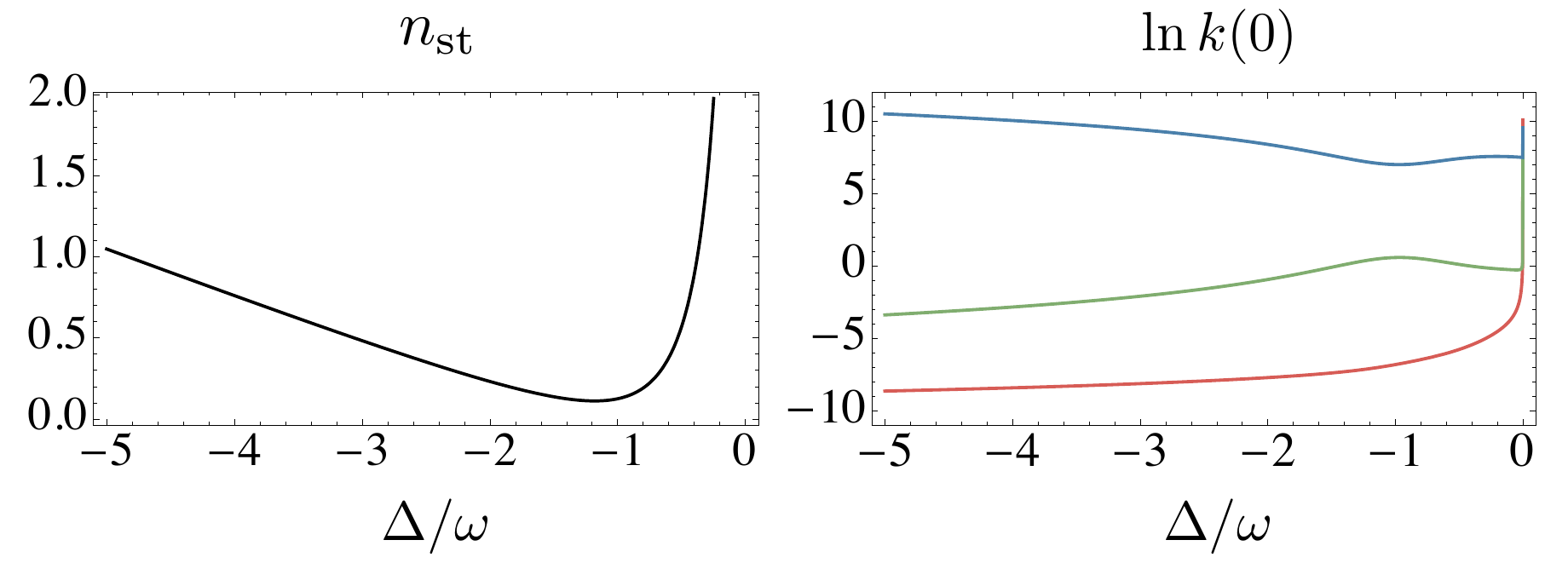}
\caption{(Color online) Left: The steady-state number of phonons, $n_\mathrm{st}$, in the vibrational state of a trapped ion against the detuning $\Delta$ between the ion's internal transition and the laser field; note that $n_\mathrm{st}$ is independent of the driving. Right: $\ln k(0)$ for the same situation, and for three different driving scenarios; from top to bottom we have (i) periodic driving with $\mathcal{F}=10$ (blue curve), (ii) constant driving with $\mathcal{F}=10$ (green), and (iii) no driving ($\mathcal{F}=0$; red). We have used the parameters $\delta=0$, $\Omega/\omega=0.5$, $\gamma/\omega=1$, $\varphi=\pi/4$.}
  \label{fig:ion}
\end{figure}

\noindent\emph{Ion/optomechanical cooling case.}---Let us consider now a trapped ion that is laser-cooled by a field in a standing-wave configuration; application to optomechanics follows analogously. In the Lamb--Dicke limit, the system can be modeled as a quantum harmonic oscillator coupled to a single bath. The corresponding coefficients $\Gamma_1$ and $\bar\Gamma_1$ can be found in Ref.~\cite{Cirac1992} and depend on the frequency of ion motion $\omega$, the Rabi frequency associated with the internal degree of freedom of the ion $\Omega$, the relative position of the ion with respect to the standing wave $\varphi$, the spontaneous emission rate of the ion, and the detuning between the laser and the ion transition frequency $\Delta$. It can be shown that the steady-state mean number of excitations of the ion motion is $n_\mathrm{st} = \bar\Gamma_1/(\Gamma_1-\bar\Gamma_1)$. A similar approach can be used to treat the sideband cooling of an optomechanical oscillator~\cite{Marquardt2007}. In scenarios such these, where we only have one bath, the counting process referring to the net number of excitations exchanged is meaningless, since it gives $\theta(s)=0$ for all $s$. We therefore modify the arguments in the preceding sections and focus on the \emph{outgoing flux} of quanta $K$. The results obtained before will take similar form, with the modification $f_\pm(s)\to \Gamma_1 (e^{-s}-1)/2 $.
From what was stated previously we see that with no driving, there is a direct relation between $n_\mathrm{st}$ and the statistics of outgoing quanta: $k(0)=2n_\mathrm{st}\Gamma_1$ and $Q(0)=2n_\mathrm{st}^2 \Gamma_1/\bar\Gamma_1$. Thanks to the large-deviation function \eref{theta}, measuring the different moments of the outgoing flux of quanta yields directly relevant physical properties of the system, e.g., the mean number of excitations.

Let us consider a particular set of parameters to illustrate our results; we take a small Rabi frequency $\Omega=\omega/2$, a negative detuning $\Delta$, and a relative position $\varphi=\pi/4$, following Ref.~\cite{Cirac1992}. The expected steady-state excitation number of the ion motion is reported against $\Delta$ in the upper panel of \fref{fig:ion}. We see that a minimum population close to zero is reached for suitable detuning. In the lower panel we report the activity, on a logarithmic scale, under the same conditions. The red line corresponds to \eref{theta} with no driving. We can see that the activity, especially for large $\lvert\Delta\rvert$, is extremely small; this makes it hard to collect enough excitations to compute the statistics of the outgoing quanta. Nevertheless, through adequate driving this activity can be enhanced, leading to conditions suitable to the experimental analysis of the statistics. In the lower panel of \fref{fig:ion}, the green curve represents the activity for a scenario with strong continuous excitation, where we see that the activity is enhanced by almost five orders of magnitude. Even more efficiently, in blue is represented the situation with driving resonant at the oscillator frequency. We see in this case an increase of about ten orders of magnitude in the rate of emitted quanta. Both situations are more favourable for determining the internal state of the ion through statistical analysis of the outgoing excitations without modifying its internal temperature. The present approach therefore allows us to easily tailor the output of our system though appropriate driving.

\noindent\emph{Conclusions.}---We have presented an analytic phase-space approach to the determination of the large-deviation function associated with the dynamics of a general, driven or undriven, quadratic bosonic system coupled to a Markovian bath. Physical quantities related to the statistics of quanta exchanged, such as the rate (activity) and the Mandel $Q$-factor, can be fully determined from the analytical expression of the large-deviation function that we have gathered. By comparing our result for a quantum harmonic oscillator coupled to two thermal baths to its classical counterpart, we have shown that at low temperatures the rate of excitations exchanged between the system and its bath is suppressed with respect to the classical case.  

We then extended this approach, applying it to a driven harmonic oscillator; in this case we obtained an analytical formulation of the large-deviation function associated with the flux of quanta. As a practical example, we looked at a trapped ion or optomechanical system, where we predicted that an appropriate driving scheme can considerably increase the mean rate of quanta emitted by the system, \emph{without} changing its internal state. This paves the way to a more effective analysis of the motional degree of freedom of the ion through the statistics of the emitted quanta. In this way, we have shown that far form being exclusively descriptive, the method proposed here can offer a means to analyse a quantum system through the excitations it exchanges with its environment.

This work was supported by the UK EPSRC (EP/L005026/1 and EP/J009776/1), the John Templeton Foundation (grant ID 43467), the EU Collaborative Project TherMiQ (Grant Agreement 618074), and the Royal Commission for the Exhibition of 1851. Part of this work was supported by COST Action MP1209 ``Thermodynamics in the quantum regime.''


\begin{thebibliography}{99}
\bibitem{Bellomo2007} B. Bellomo, R. Lo Franco, and G. Compagno, \href{http://dx.doi.org/10.1103/PhysRevLett.99.160502}{Phys. Rev. Lett. {\bf 99}, 160502 (2007).}
\bibitem{Gardiner2010} C. W. Gardiner, and P. Zoller, {\it Quantum Noise} (Springer, Berlin, Heidelberg, 2004).
\bibitem{Belzig2001} W. Belzig and Yu. V. Nazarov, \href{http://dx.doi.org/10.1103/PhysRevLett.87.197006}{Phys. Rev. Lett. {\bf 87}, 197006 (2001).}
\bibitem{Garrahan2010} J. P. Garrahan and I. Lesanovsky, \href{http://dx.doi.org/10.1103/PhysRevLett.104.160601}{Phys. Rev. Lett. {\bf 104}, 160601 (2010).}
\bibitem{Lecomte2005} V. Lecomte, C. Appert-Rolland, and F. van Wijland, \href{http://dx.doi.org/10.1103/PhysRevLett.95.010601}{Phys. Rev. Lett. {\bf 95}, 010601 (2005).}
\bibitem{Marconi2008} U. Marconi, A. Puglisi, L. Rondoni, and A. Vulpiani, \href{http://dx.doi.org/10.1016/j.physrep.2008.02.002}{Phys. Rep. {\bf 461}, 111 (2008).}
\bibitem{Touchette2009} H. Touchette, \href{http://dx.doi.org/10.1016/j.physrep.2009.05.002}{Phys. Rep. {\bf 478}, 1 (2009).}
\bibitem{Ates2012} C. Ates, B. Olmos, J. P. Garrahan, and I. Lesanovsky, \href{http://dx.doi.org/10.1103/PhysRevA.85.043620}{Phys. Rev. A {\bf 85}, 043620 (2012).}
\bibitem{Genway2014} S. Genway, W. Li, C. Ates, B. P. Lanyon, and I. Lesanovsky, \href{http://dx.doi.org/10.1103/PhysRevLett.112.023603}{Phys. Rev. Lett. {\bf 112}, 023603 (2014).}
\bibitem{Pigeon2014} S. Pigeon, A. Xuereb, I. Lesanovsky, J. P. Garrahan, G. De Chiara, and M. Paternostro, \href{http://www.arxiv.org/abs/1409.0422}{arXiv:1409.0422.}
\bibitem{Cahill1969} K. E. Cahill and R. J. Glauber, \href{http://dx.doi.org/10.1103/PhysRev.177.1857}{Phys. Rev. {\bf 177}, 1857 (1969).}
\bibitem{Mandel1979} L. Mandel, \href{http://dx.doi.org/10.1364/OL.4.000205}{Opt. Lett. {\bf 4}, 205 (1979).}
\bibitem{NoteQfactor} For a reference bath with $\bar{\Gamma }_1>0$, e.g., a thermal bath at nonzero temperature, we must redefine the counting process to only account for the \emph{outgoing} flux. For the associated $Q$-factor, we find the same expression as the case of a bath with $\bar{\Gamma}=0$, thus allowing us to conclude that no anti-bunching of the outgoing flux can be observed for a quantum harmonic oscillator coupled to multiple baths.
\bibitem{Lebowitz2011} J. L. Lebowitz, H. Spohn, \href{http://dx.doi.org/10.1023/A:1004589714161}{J. Stat. Phys. {\bf 95}, 333 (1999).}
\bibitem{Visco2006} P. Visco, \href{http://dx.doi.org/10.1088/1742-5468/2006/06/P06006}{J. Stat. Mech. {\bf 2006}, P06006 (2006).}
\bibitem{Fogedby2011} H. C. Fogedby, and A. Imparato, \href{http://dx.doi.org/10.1088/1742-5468/2011/05/P05015}{J. Stat. Mech. {\bf 2011}, P05015 (2011).}
\bibitem{NoteLimits} In other words, $\lim_{\delta\to0}\theta_\mathrm{d}(s)\neq\theta_\mathrm{d}(s)\rvert_{\delta=0}$. The reason for this is that $\lim_{t\to\infty}\int\cos(\delta\tau)\rm{d}\tau/t=0$ for any nonzero $\delta$, so the limits $t\to\infty$ and $\delta\to0$ do not commute.
\bibitem{Cirac1992} J. I. Cirac, R. Blatt, P. Zoller, and W. D. Phillips, \href{http://dx.doi.org/10.1103/PhysRevA.46.2668}{Phys. Rev. A {\bf 46}, 2668 (1992).}
\bibitem{Marquardt2007} F. Marquardt, J. P. Chen, A. A. Clerk, and S. M. Girvin, \href{http://dx.doi.org/10.1103/PhysRevLett.99.093902}{Phys. Rev. Lett. {\bf 99}, 093902 (2007).}

\end{thebibliography}
\end{document}